\documentclass[12pt,oneside]{amsart}
\usepackage[margin=1in]{geometry}
\usepackage{amsmath,amsthm,amssymb,amsfonts}
\usepackage{graphicx}
\usepackage{latexsym}
\usepackage[T2A]{fontenc}
\usepackage[utf8]{inputenc}
\usepackage[russian,english]{babel}
\usepackage{natbib}
\usepackage{ragged2e}
\usepackage{setspace}
\usepackage{xcolor}
\usepackage{csquotes}

\theoremstyle{definition}

\begin{document}
\onehalfspacing

\thispagestyle{empty}

\title{Minimal Envy and Popular Matchings}

\author{Aleksei Y. Kondratev}

\author{Alexander S. Nesterov}

\thanks{\phantom{ } Aleksei Y. Kondratev --- National Research University Higher School of Economics, 16, Soyuza Pechatnikov st., St.~Petersburg, 190121, Russia; Institute for Problems of Regional Economics RAS, 38, Serpuhovskaya st., St.~Petersburg, 190013, Russia --- akondratev@hse.ru --- http://orcid.org/0000-0002-8424-8198\\
\phantom{ } Alexander S. Nesterov --- National Research University Higher School of Economics, 16, Soyuza Pechatnikov st., St.~Petersburg, 190121, Russia --- asnesterov@hse.ru --- http://orcid.org/0000-0002-9143-2938\\
\phantom{ } Support from the Basic Research Program of the National Research University Higher School of Economics is gratefully acknowledged. 
\\
\phantom{ } Acknowledgements will be added in the final version of the paper.\\
\phantom{ } The preliminary version of this paper has circulated under the title ``Random Paths To Popularity In Two-Sided Matching''.}

\date{08 February 2019}

\begin{abstract}
We study ex-post fairness in the object allocation problem where objects are valuable and commonly owned. A matching is fair from individual perspective if it has only inevitable envy towards agents who received most preferred objects -- \textit{minimal envy matching}. A matching is fair from social perspective if it is supported by majority against any other matching -- \textit{popular matching}. Surprisingly, the two perspectives give the same outcome: when a popular matching exists it is equivalent to a minimal envy matching. 

We show the equivalence between global and local popularity: a matching is popular if and only if there does not exist a group of size up to 3 agents that decides to exchange their objects by majority, keeping the remaining matching fixed. We algorithmically show that an arbitrary matching is path-connected to a popular matching where along the path groups of up to 3 agents exchange their objects by majority. A market where random groups exchange objects by majority converges to a popular matching given such matching exists. 

When popular matching might not exist we define \textit{most popular matching} as a matching that is popular among the largest subset of agents. We show that each minimal envy matching is a most popular matching and propose a polynomial-time algorithm to find them. 


\medskip
\medskip


\medskip

\end{abstract}

\maketitle

\thispagestyle{empty}

\newpage

\section{Introduction}\nopagebreak

Consider a situation where a group of agents is being allocated a set of valuable indivisible objects, such that each agent gets at most one object. Initially the objects are commonly owned and each agent has the same right to each object. In practice, such situations include social housing programs, assigning children to daycare and to primary schools, distributing transplants to patients, and so forth. In each of these situations fairness is one of the primary concerns, but the very fact that objects are valuable, indivisible and commonly owned often precludes existence of an ex-post fair allocation.\footnote{When objects are not valuable and can be wasted \cite{KestenYazici12} propose an ex-post envy-free strategy-proof rule that Pareto-dominates all such rules. For the case when objects are privately owned or have priorities over agents and we only need to eliminate ``justified envy'' see the prolific school choice literature stemming from the seminal paper by \cite{AbdulkadirogluSonmez03}. For the case when objects can be goods and bads and agents might receive more than one object see the fair division literature, e.g. \cite{BogomolnaiaMoulinSandomirskiyYanovskaya17}.} For an illustration consider Example~1 below.

\medskip

\textbf{Example 1.} Let there be four agents $I=\{1,2,3,4\}$ that
have the following preferences (Table~1) over four houses
$H=\{a,b,c,d\}$:

{
\singlespacing
\begin{center}
Table 1. Preference profile and matching $\mu$ \\* \vskip 1mm
\begin{tabular}{|c|c|c|c|}
\hline
1 & 2 & 3 & 4   \\
\hline
a & \underline{a} & a & a \\
\underline{d} & b & \underline{b} & \underline{c} \\
c & d & c & b \\
b & c & d & d \\
\hline
\end{tabular}
\end{center}
}
\medskip

What would be a fair allocation in such setting? There are two perspectives to judge if an allocation is fair or not: individual perspective -- we want each agent to be treated fairly compared to other agents, and a social perspective -- we want a matching to be preferred by the agents collectively. 

From the individual perspective the most common approach is to eliminate envy: to find an allocation such that no agent envies another agent.\footnote{The two other standard approaches -- symmetry and fair-share-guarantee -- are useful for the ex-ante consideration yet hardly applicable ex-post.} Yet eliminating envy completely is rarely feasible. In Example 1, for each matching there will always be exactly three agents who envy the agent that receives object $a$. Such envy is generally \textit{inevitable}: in each matching an object causes envy among those agents who top-ranked it towards the agent who got it.

But if we exclude this inevitable envy towards top-ranked objects from consideration, we might be able to eliminate the remaining envy. Consider matching $\mu=(1d,2a,3b,4c)$. When we exclude inevitable envy towards agent 2 caused by top-ranked object $a$, then agents 1,3,4 do not envy each other. A matching that minimizes the number of agents who have envy modulo inevitable envy caused by top-ranked objects we call a \textbf{minimal envy matching}.

In general, in each minimal envy matching there are two groups of agents: in the first group each agent gets his top-ranked object and thus does not envy (agent 2 in matching $\mu$ in our example), and the second group containing all other agents. The second group has envy towards the first group and possibly also among each other (agents 1,3,4 in our example). Envy towards the first group is inevitable and is of fixed size as we only have a fixed number of top-ranked objects to allocate.\footnote{For a matching to be minimal-envy, each top-ranked object must be given to one of the agents that top-ranked it since otherwise we can reduce the number of envying agents at least by 1.}
In the second group each agent envies at least one agent in the first group. Yet, envy within the second group might be manageable. If it happens that no agent in the second group has envy towards another agent in the second group, then such minimal-envy matching is a natural candidate to be called an individually fair matching.

From the social perspective we let the agents compare matchings collectively and put the decision to a popular vote. If it happens that there is a matching that is supported by majority against any other matching, then this matching is also a natural candidate to be called a socially fair matching. In other words this matching is a weak Condorcet winner in a voting problem where each matching is a candidate and each agent is a voter. Such matching is commonly referred to as a \textbf{popular matching}. In our example the same matching  $\mu=(1d,2a,3b,4c)$ is a popular matching.

Perhaps surprisingly, the individual perspective and the social perspective always lead to the same conclusion: whenever a popular matching exists it is equivalent to the minimal envy matching (Theorem~1 and Observation~1). Motivated by this equivalence we further study popular and minimal envy matchings and show the following three main results.

First we show that a matching is popular globally if and only if it is popular locally (Theorem~2). Specifically, for a matching $\mu$ to be popular for the entire set of agents it is necessary and sufficient that it is popular for an arbitrary triple of agents. That is, for each triple of agents, keeping the matching of the other agents fixed, there does not exist a matching $\mu'$ that is supported by majority (or, equivalently, has fewer envying agents) compared to the original matching $\mu$. In particular, this triple of agents does not want to swap their objects based on majority.

This result has two other interesting interpretations. From the cooperative standpoint, if for each coalition of agents we take the majority rule as the solution concept, then for a matching to be in the core it is enough to check coalitions of size up to three. The analogous result for the marriage market states that the set of pairwise stable matchings coincides with the core \citep{RothSotomayor89}. 

From the axiomatic standpoint, our result means that a correspondence that induces popular matchings satisfies consistency (an axiom requiring that a solution to a larger problem coincides with a solution to all ``reduced'' problems -- where some agents ``leave'' with their matching) and converse consistency (an axiom requiring that knowing the solution to all reduced problems is sufficient to restore the solution of the larger problem).\footnote{For a review of consistency principle behind these two axioms see \cite{Thomson11}.}




Our second result (Theorem~3) shows that by means of these majority-based exchanges in triples of agents we can always reach a popular matching, given such a matching exists. In other words, each matching is path-connected with a popular matching where along the path each next matching is ``more popular'' for a specific triple of agents, while the matching of the other agents is fixed. 

The proof is constructive: we provide an algorithm to reach a popular  matching. We start with an arbitrary matching, find a specific triple of agents that modify their matching keeping the matching of the other agents fixed, then we find another triple and so on until we reach a matching where no such triple of agents exists. (In case we end up in a loop we conclude that a popular matching does not exist.) Despite being greedy our algorithm is computationally efficient. The speed of the algorithm is quadratic in the number of agents, as is the speed of the original algorithm by \cite{Abraham07}.\footnote{\cite{Abraham07} propose an $O(n+m')$ algorithm, where $m'$ is the total length of all preferences, i.e. up to $m'=|I|\cdot |H|$, where $n=|I|$ is the number of agents and $|H|$ is the number of houses.}

The result above has a peculiar implication: arbitrary local majority-based exchanges as in our algorithm lead to a globally popular 
matching. Imagine a decentralized market where agents meet at random in arbitrary groups and exchange their objects based on majority of this group. Corollary~1 postulates that this market eventually converges to a popular matching whenever such matching exists.

This finding is analogous to the result by \cite{RothVate90} about
convergence in a marriage market. There, one matching is modified
locally by a random blocking pair of a man and a woman that prefer each other
over their current matches. As this man and this woman match, their
previous partners become unmatched, and these changes constitute a
new matching. Then a new blocking pair is considered, a new matching
is formed, and so forth, and \cite{RothVate90} show that a random sequence of
these matchings leads to a stable matching. This result also holds in the setting of many-to-many matching \citep{KojimaUnver08}.



Our concept of a fair matching above as well as the results so far rely on that a popular matching exists, which is often not the case. Our third main result is a constructive extension of popular matchings to the cases where a popular matching does not exist. We say that a matching is \textbf{most popular} if it is popular among the largest number of agents. This is one of the standard extensions of the Condorcet voting rule due to \cite{Young77}. 

To illustrate, let us modify Example 1 by adding agent 2' with preferences identical to agent 2. In this case a popular matching no longer exists. However, $\mu$ remains the most popular matching as we only need to exclude agent 2' to restore popularity. Alternatively, we could get a popular matching by excluding agent 2 or agent 3.

We show that most popular matchings are related to minimal envy matchings: each minimal envy matching is a most popular matching (Theorem~4).

We then provide a polynomial-time algorithm to find minimal envy matchings and show that the outcomes of this algorithm are Pareto efficient and coincide with popular matchings whenever those exist (Theorem~5). The fact that the algorithm is polynomial-time is, perhaps, surprising given that the Young's rule is NP-hard.\footnote{More precisely, it is  $P^{NP}_{||}$-complete as shown by \cite{RotheSpakowskiVogel03}, whereas the speed of our algorithm is cubic in the number of agents $|I|$.} To the best of our knowledge this is the first computationally feasible extension of popular matchings in the object allocation problem (see subsection 1.1 for details). 

The remainder of the paper is organized as follows. This section ends with a brief review on popular matchings and fairness criteria in object allocation problem. Section 2 presents the model, section 3 presents the characterization result, section 4 presents the algorithm to find popular matchings and the convergence result, section 5 presents the extension of popularity to the cases where popular matchings do not exist and shows that this extension is consistent with minimal envy.

\subsection{Background on fairness and popularity in object allocation problem}\nopagebreak

The object allocation problem where agents exchange indivisible objects
(referred to as houses) without money was first introduced by \cite{ShapleyScarf74}, the
assignment problem where all houses are initially commonly owned was first
studied by \cite{HyllandZeckhauser79}. 

The literature on ex-ante fairness is prolific (see, e.g. \cite{Nesterov17} and references therein, and recent surveys by \cite{BouveretChevaleyreMaudet16} and \cite{Moulin18fair}). In contrast, the literature on ex-post fairness that is a focus of this paper is extremely scarce because the standard approaches face fundamental difficulties. The symmetry approach (e.g. equal treatment of equals: agents with identical preferences should receive identical outcomes) necessarily leads to waste of valuable resources. The approach called fair share guaranteed (e.g. equal division lower bound: each agent should prefer his outcome to equal division of resources) is hard to conceptualize ex-post since there is no obvious candidate to be called the fair share. The only remaining standard approach is to eliminate envy between agents \citep{Foley67}, yet this elimination also leads to wastes \citep{KestenYazici12}.

The concept of popularity was first
introduced by \cite{Gardenfors75} for the marriage problem
\citep{GaleShapley62}, where popularity coincides with stability, and was
applied to house allocation problem only recently by
\cite{Abraham07}. The characterization by \cite{Abraham07} also allows non-strict preferences.\footnote{This setting was further
generalized to the case with ties and matroid constraints by
\cite{Kamiyama17} and to the case with two-sided preferences and
one-sided ties by \cite{CsehHuangKavitha17}. The many-to-one matching problem, where each house has a capacity was studied by \cite{SngManlove10} and \cite{KavithaNasre11}, and the many-to-many problem was studied by \cite{Paluch14}.}

Among other properties of popular matchings the literature extensively studied existence and multiplicity. \cite{Mahdian06} shows that a popular matching is likely to exist whenever preferences are uniformly random and the number of houses is approximately 1.42 times larger than the number of agents. To restore existence \cite{McCutchen08} proposes least-unpopularity criteria to find the ``most'' popular matching; finding his least-unpopular matchings is, in general, NP-hard.

Another way to ensure popularity is to consider mixed matchings,
i.e. lotteries over matchings, and a straightforward generalization
of the popularity property; \cite{KavithaMestreNasre11} show that a popular
mixed matching always exists and propose an efficient algorithm to
find one. The recent literature has studied compatibility of popularity in mixed matchings with various fairness and incentive properties \citep{AzizBrandtStursberg13,BrandtHofbauerSuderland17}.

The problem of counting the number of popular matchings has been
addressed by \cite{McDermidIrving11} for the case of strict preferences
and by \cite{Nasre14} and \cite{Acharyya14} for the case of weak
preferences. For the case of large number of popular matchings \cite{KavithaNasre09} propose ways to find popular matchings that are optimal in terms of envy and rank distribution.
Popularity with agents having different weights has been studied by \cite{Mestre14}.

Related to our convergence result, \cite{AbrahamKavitha10} consider the
popularity-improvement paths from an arbitrary matching. The main
finding is that, given a popular matching exists, it can be attained
by at most two steps using an efficient algorithm.

For a recent review on popular matchings see \cite{Cseh17} and \cite{KlausManloveRossi16}.

\section{The Model}\nopagebreak

Let $I$ be a set of agents and $H$ be a possibly larger set of houses, $|H|\geq |I|$. Each agent $i\in I$ is endowed with a strict preference relation $\succ_i$ over the set of houses $H\cup\{\emptyset\}$ (i.e. $\succ_i$ is a linear order), and $i$ prefers each house $h\in H$ over having no house, $h\succ_i \emptyset$.\footnote{For each agent~$i$, having no house $\emptyset$ can be considered as his last resort (fictitious house) $l_i$. All results remain true when agents have short preference lists with last resort and/or the set of houses is smaller than than the set of agents.} The collection of individual preferences of all agents $\succ=(\succ_i)_{i\in I}$ is referred to as the \textbf{preference profile}. A~triple $(I,H,\succ)$ constitutes a \textbf{problem} (aka the object allocation problem, house allocation problem, assignment problem and two-sided matching problem with one-sided preferences).

A solution to a problem is a \textbf{matching} $\mu$ -- a mapping from $I\cup H$ on $I\cup H \cup {\emptyset}$. By definition agent $i\in I$ is said to be matched to house $h\in H$ in matching $\mu$ if $\mu(i)=h$ and also $\mu(h)=i$. If some agent or house remain unmatched, we say that they are matched to~$\emptyset$.

For an illustration consider Example~2 below.

\smallskip

\textbf{Example 2.} Let there be four agents $I=\{1,2,3,4\}$ that have the following preferences (Table~2) over four houses $H=\{a,b,c,d\}$:

{
\singlespacing
\begin{center}
Table 2. Preference profile and matching $\mu$ \\* \vskip 2mm
\begin{tabular}{|c|c|c|c|}
\hline
1 & 2 & 3 & 4   \\
\hline
\underline{a} & d  & a & \underline{d} \\
d & \underline{b} & \underline{c} & b \\
b & a & b & c \\
c & c & d & a \\
\hline
\end{tabular}
\end{center}
}
\smallskip

For any two matchings $\mu,\mu$ and a subset of agents $J\subset I$ define \textbf{pairwise comparison} $PC_J(\mu,\mu')$ as the number of agents in $J$ that prefer their house in $\mu$ over their house in~$\mu'$,
\begin{equation*}
    PC_J(\mu,\mu') = |\{j \in J : \mu(j) \succ_j \mu'(j) \}|.
\end{equation*}

A matching $\mu$ is called \textbf{popular} if there does not exist another matching $\mu'$ such that $\mu'$ is preferred over $\mu$ by majority within entire
set of agents $I$: $PC_I(\mu',\mu)>PC_I(\mu,\mu')$.

In the profile from Table~2 consider matchings $\mu = (1a,2b,3c,4d)$ and $\mu' = (1d,2c,3a,4b)$. Agents~1,2,4 prefer $\mu$ over $\mu'$ and agent~3 prefers $\mu'$ over $\mu$. Hence $PC_I(\mu,\mu') = 3 > 1 = PC_I(\mu',\mu)$ and matching $\mu'$ is not popular.

The following simple terminology is useful in order to discuss popular matchings. For each agent $i\in I$ let us call his most preferred house in $H$ as \textbf{$i$'s first house}: $FH(i)=h$ such that for each $h'\in H$ and $h'\neq h$ it holds that $h\succ_i h'$. \textbf{The set of all first houses} is denoted as $FH=\{FH(i)\}_{i\in I}$. For each house $h$ let us call agents for whom $h$ is the first house as \textbf{$h$'s first agents}: $FA(h)=\{i\in I : h = FH(i)\}$.

For each agent $i$ let us call his most preferred house among all non-first houses as \textbf{$i$'s second house}: $SH(i)=h$ such that $h\in H\setminus FH$, for each $h'\in H\setminus FH$ and $h'\neq h$ it holds that $h\succ_i h'$. \textbf{The set of all second houses} is denoted as $SH=\{SH(i)\}_{i\in I}$. Note that sets $FH$ and $SH$ are disjoint, i.e. no agent's second house can be a first house for any other agent. For each house $h$ let us call agents for whom $h$ is the second house as \textbf{$h$'s second agents}: $SA(h)=\{i\in I : h = SH(i)\}$. Each house $h$ that is worse than agent $i$'s second house, $SH(i)\succ_i h$, is \textbf{$i$'s bad house}. 

The next theorem is the original characterization of popular matchings.

\smallskip

\noindent \textbf{Theorem 1} \citep{Abraham07}. For a problem $(I,H,\succ)$, a matching $\mu$ is popular if and only if the following two conditions hold:
\begin{enumerate}
    \item Each first house $f\in FH$ is matched to one of its first agents, $\mu(f)\in FA(f)$;
    \item Each agent $i\in I$ gets either his first house or his second house, $\mu(i)\in FH(i)\cup SH(i)$. 
\end{enumerate}

\smallskip

In the profile from Table~2, the set of first houses is $FH=\{a,d\}$ and the set of second houses is $SH=\{b,c\}$. Hence,  there are only two popular matchings $\mu=(1a,2b,3c,4d)$ and $(1a,2d,3c,4b)$. Each other feasible matching either assigns some agent his bad house, or does not distribute first houses among agents that prefer them most (or both, as in matching $\mu'=(1d,2c,3a,4b)$ where agent 2 gets his bad house $c$ and the first house $d$ is not assigned to agents 2 or 4 that value it the most).

For a problem $(I,H,\succ)$ and a matching $\mu$, the \textbf{set of envying agents} $E(\mu,I,H,\succ)$ contains each agent $i$ for whom there is a house $h$ that $i$ prefers over his matching:
\begin{equation*}
    E(\mu,I,H,\succ)=\{i\in I: \quad \mbox{there exists} \quad h \in H \quad \mbox{such that} \quad h\succ_i \mu(i)\}.
\end{equation*}

A matching $\mu$ is a \textbf{minimal envy matching} for problem $(I,H,\succ)$ if it:
\begin{enumerate}
    \item minimizes the number of envying agents $|E(\mu,I,H,\succ)|$ (inevitable envy), and among such matchings also
    \item minimizes the number of envying agents $|E(\mu,I',H',\succ_{I',H'})|$ in the reduced problem (remaining envy), where $I'=E(\mu,I,H,\succ),H'=H\setminus \mu(I\setminus E(\mu,I,H,\succ))$.
\end{enumerate}

In the profile from Table~2, the first part of this definition requires that house $a$ is given to either agent 1 or agent 3, and that house $d$ is given to either agent 2 or agent~4. The second part of the definition requires that envy among the remaining agents is minimized, which is achieved, for example, in the same matching $\mu=(1a,2b,3c,4d)$. 

We conclude this section by formally stating the equivalence between popular and minimal envy matchings.

\medskip

\noindent {\bf Observation 1}.\label{obs-envy-pop} Whenever a popular matching exists, the set of minimal envy matchings coincides with the set of popular matchings.

\medskip

We leave the proof of Observation 1 till section 5 where we extend the definition of popular matchings such that the equivalence similar to Observation~1 holds even in cases when a popular matching does not exist (see Observation~2).

\section{Characterization of Popular Matchings}\nopagebreak

In this section, we introduce a new characterization of popular matchings using their local properties.

For an illustration consider the profile from Table~2 and popular matching $\mu = (1a,2b,3c,4d)$. We see that in each triple of agents, when we only consider the houses owned by this triple, each such (reduced) matching is popular within this triple. The following Table~3 illustrates the popularity within each triple. To check that each reduced matching is popular, recall the characterization in Theorem~1: each first house is given to one of its first agents, and each agent gets either his first house or his second house.

{
\singlespacing
\begin{center}
Table 3. Popular matching $\mu$ reduced to each possible triple of agents \\* \vskip 2mm
\begin{tabular}{|c|c|c|}
\hline
1 & 3 & 4  \\
\hline
\underline{a} & a & \underline{d} \\
d & \underline{c} & c \\
c & d & a \\
\hline
\end{tabular}
\quad
\begin{tabular}{|c|c|c|}
\hline
1 & 2 & 3  \\
\hline
\underline{a} & \underline{b} & a \\
b & a & \underline{c} \\
c & c & b \\
\hline
\end{tabular}
\quad
\begin{tabular}{|c|c|c|}
\hline
1 & 2 & 4  \\
\hline
\underline{a} & d & \underline{d} \\
d & \underline{b} & b \\
b & a & a \\
\hline
\end{tabular}
\quad
\begin{tabular}{|c|c|c|}
\hline
2 & 3 & 4  \\
\hline
d & \underline{c} & \underline{d} \\
\underline{b} & b & b  \\
c & d & c \\
\hline
\end{tabular}
\end{center}
}

\medskip

Let us formally define this property. For a problem $(I,H,\succ)$, we say that a matching $\mu$ is \textbf{locally popular} if for each three agents $i,j,k\in I$ there does not exist a matching $\mu'$ same as $\mu$ for each other agent $i'\notin \{i,j,k\}$, $\mu'(i')=\mu(i')$, and such that $\mu'$ is preferred over $\mu$ by majority within this triple of agents, $PC_{\{i,j,k\}}(\mu',\mu)>PC_{\{i,j,k\}}(\mu,\mu')$. In other words, (reduced) matching $\mu$ is popular in the reduced problem $(I',H',\succ_{I',H'})$ where $I' = \{i,j,k\}$ and $H' = H\setminus \mu(I\setminus \{i,j,k\})$. 

We arrive to our first main result: the equivalence between global popularity and local popularity.

\medskip

\noindent {\bf Theorem 2}.\label{th_local-pop} A matching is popular if and only if it is locally popular.

\medskip

The proof of Theorem~2 is in the appendix. As an immediate corollary we get the characterization of popular matchings by \cite{Abraham07}, here we provide an illustrative proof.

\medskip
\textsl{Proof of Theorem 1}. The ``if'' part is straightforward since it is
enough to check only triples of agents. In each such triple only some agent $i$ with a second house can become better off, but each better
house $f\succ_i SH(i)$ is already matched to one of its first agents
$j=\mu(f)\in FA(f)$, and making $i$ better off requires making $j$ worse off, which cannot be supported by majority.

We prove the ``only if'' part by contradiction. Let condition (1) of Theorem~1 be violated: some first house $f$ is not allocated to one of its first agents, $\mu(f)\notin FA(f)$. Then each $f$'s first agent $i\in FA(f)$, the owner of $f$ agent $j=\mu(f)$ and the owner of $j$'s first house agent $k=\mu(FH(j))$ form a triple for which $\mu$ is not popular. (If $\mu(f)=\emptyset$ then we can choose any $j,k$; and if $\mu(FH(j))\in \{\emptyset,i\}$ then we can choose any $k$.)


Let condition (2) of Theorem~1 be violated: some agent $i_1$ gets a bad house $t$ in matching~$\mu$. Then there is a triple of agent $i_1$, the owner of $i_1$'s second house agent $i_2=\mu(SH(i_1))$, and the owner of $i_2$'s first house agent $i_3=\mu(FH(i_2))$ for whom $\mu$ is not popular. (If $\mu(SH(i_1))=\emptyset$ then we can choose any $i_2,i_3$.) \hfill
$\blacksquare$

\section{The Algorithm and Random Paths to Popular Matchings}\nopagebreak


Here we present a new algorithm to find a popular matching. Specifically, for each matching $\mu$ we consider all ``neighbouring'' matchings $\mu'$ that is matchings where at most three agents are matched to different house than in $\mu$. If one, two, or three agents are matched differently in $\mu$ and $\mu'$, and more than half of these agents prefer $\mu'$ over $\mu$, then we say that $\mu$ and $\mu'$ are connected by a \textbf{local popular exchange}. The algorithm begins with an arbitrary matching $\mu_0$ (e.g., all agents and houses may be unmatched) and then modifies it using local popular exchanges among one, two or three agents. 

The algorithm has two parts. In the first part the algorithm assigns each first house to one of its first agents. In each current matching $\mu_j$, some agent $i_1$ starts a local popular exchange if his first house $FH(i_1)\neq \mu_j(i_1)$ belongs to some agent $i_2=\mu_j(FH(i_1))$ for whom it's not the first house, $FH(i_2)\neq \mu_j(i_2)=FH(i_1)$. In the new matching $\mu_{j+1}$, agent $i_1$ gets his first house, $\mu_{j+1}(i_1)=FH(i_1)$, agent $i_2$ also gets his first house, $\mu_{j+1}(i_2)=FH(i_2)$, the owner of that latter house, agent $i_3=\mu_j(FH(i_2))$, gets the leftover house previously owned by $i_1$, $\mu_{j+1}(i_3)=\mu_j(i_1)$. After all first houses are distributed to its first agents, each agent gets either his first house, his second house, or a bad house.

In the second part the algorithm constructs subsequences of local popular exchanges such that each subsequence decreases the number of agents with bad houses. In each current matching $\mu_j$, an agent $i_1$ who has a bad house $t=\mu_j(i_1)$ gets his second house, $\mu_{j+1}(i_1)=SH(i_1)$, the owner of that house, agent $i_2=\mu_j(SH(i_1))$, gets his first house, $\mu_{j+1}(i_2)=FH(i_2)$, the owner of that latter house, agent $i_3=\mu_j(FH(i_2))$,  gets the leftover house $t$, $\mu_{j+1}(i_3)=t$. If house $t$ is bad for agent $i_3$, then this agent continues the subsequence of exchanges; otherwise, another agent with a bad house starts a new subsequence of exchanges. If a subsequence of exchanges ends up in a loop, then there is no popular matching.

We place all technical details of the algorithm in the proof of Theorem 3 in the appendix.


Here, we illustrate how the algorithm works using the preference profile from Table~2. Let the initial matching be $\mu_0=(1b,2c,3d,4a)$. We now show how the first part of the algorithm turns $\mu_0$ into $\mu_1$ pictured below (Table~4), and the second part of the algorithm turns $\mu_1$ sequentially into~$\mu_2$, then into~$\mu_3$, then into~$\mu_4$, which is a popular matching.

{
\singlespacing
\begin{center}
Table 4. Initial matching $\mu_0$, intermediate matchings $\mu_1,\mu_2,\mu_3$, and final matching $\mu_4$. \\* \vskip 2mm 
\begin{tabular}{|c|c|c|c|}
\hline
1 & 2 & 3 & 4   \\
\hline
a & d & a & d \\
d & b & c & b \\
\underline{b} & a & b & c \\
c & \underline{c} & \underline{d} & \underline{a} \\
\hline
\end{tabular}
\quad
\begin{tabular}{|c|c|c|c|}
\hline
1 & 2 & 3 & 4   \\
\hline
\underline{a} & d & a & \underline{d} \\
d & b & c & b \\
b & a & \underline{b} & c \\
c & \underline{c} & d & a \\
\hline
\end{tabular}
\quad
\begin{tabular}{|c|c|c|c|}
\hline
1 & 2 & 3 & 4   \\
\hline
a & d & \underline{a} & \underline{d} \\
d & \underline{b} & c & b \\
b & a & b & c \\
\underline{c} & c & d & a \\
\hline
\end{tabular}
\quad
\begin{tabular}{|c|c|c|c|}
\hline
1 & 2 & 3 & 4   \\
\hline
a & \underline{d} & \underline{a} & d \\
d & b & c & b \\
\underline{b} & a & b & \underline{c} \\
c & c & d & a \\
\hline
\end{tabular}
\quad
\begin{tabular}{|c|c|c|c|}
\hline
1 & 2 & 3 & 4   \\
\hline
\underline{a} & \underline{d} & a & d \\
d & b & \underline{c} & \underline{b} \\
b & a & b & c \\
c & c & d & a \\
\hline
\end{tabular}
\end{center}
}

\medskip

\emph{First part of the algorithm.} Agent~1 owns house~$b$, his first house is $a$ owned by agent~4 whose first house is $d$. We implement the following exchange: agent~1 gets house $a$, agent~4 gets house~$d$, and house $d$'s previous owner agent~3 gets the leftover house $b$. In the resulting matching $\mu_1=(1a,2c,3b,4d)$ all first houses are assigned to agents that prefer them the most and thus the first part of the algorithm is completed.

\emph{Second part of the algorithm.} After the first part agents 1 and 4 get their first houses, agents 2 and 3 get bad houses. We take an arbitrary agent with a \emph{bad} house, e.g. agent 2 with his bad house $c$, and give him his \emph{second} house, $b$, and the owner of house $b$, agent 3, gets his \emph{first} house~$a$. The previous owner of house $a$ agent 1 gets the leftover house $c$, which is his bad house.\footnote{If we selected agent 3 instead of agent 2, the procedure would have stopped after one local popular exchange.}

In the resulting matching $\mu_2=(1c,2b,3a,4d)$ we continue with agent 1 because he got his bad house $c$ after the previous exchange. Again, we give agent 1 his second house $b$, the owner of $b$, agent 2, gets his first house $d$. The previous owner of $d$, agent 4, gets the leftover house $c$, which is his bad house. 

In the resulting matching $\mu_3=(1b,2d,3a,4c)$ we continue with agent 4 because he got his bad house $c$ after a previous exchange. Agent 4 gets his second house $b$, the owner of $b$, agent 1 gets his first house $a$, and the owner of $a$, agent 3, gets the leftover house $c$. House $c$ is agent 3's second house and there is no more agent with a bad house. We arrived to a popular matching $\mu_4=(1a,2d,3c,4b)$. 

\medskip

Next we present our second main result: local popular exchanges always lead to a popular matching, whenever such matching exists.


\medskip

\noindent {\bf Theorem 3}. Let $\mu_0$ be an arbitrary matching for a problem $(I,H,\succ)$ and let a popular matching exist. Then there exists a finite sequence of matchings $\mu_0, \mu_1, \ldots, \mu_l$ connected by local popular exchanges such that $\mu_l$ is popular, and $l\leq (|I|^2-|I|+2)/2$.

\medskip

The constructive proof of Theorem~3 is based on the algorithm above and it is placed into the appendix. The algorithm finds a popular matching or verifies that a popular matching does not exist in quadratic time in the number of agents $|I|$. Since our algorithm is finite we immediately get the convergence result for the corresponding decentralized market.

We represent the sequence of matchings as a finite Markov chain. The set space is the set of all matchings. The transition probabilities between the states depend on how many agents become better off or worse off in one state compared to the other.  The transition probability is positive for all local popular exchanges, otherwise the transition probability is zero.

\medskip

\noindent {\bf Corollary 1.} For any initial matching, the random sequence of local popular exchanges converges with probability one to a popular matching whenever such matching exists.

\medskip

The restriction to the groups of up to three agents is not compulsory as the same algorithm works when groups of larger size are also allowed. The convergence result also holds for popular exchanges of arbitrary sizes.




The original result in \cite{RothVate90} was partially motivated by the example in \cite{Knuth76} where he shows that a sequence of blocking pairs might have an infinite cycle and might never converge to stability. The same is true in our setting: even when a popular matching exists, the sequence of (local) popular exchanges might have cycles. To see that let us consider a preference profile and a cycle (Table~5).

{
\singlespacing
\begin{center}
Table 5. Cycle with 4 matchings: $\mu_1,\mu_2,\mu_3,\mu_4$ \\*
\vskip 2mm
\begin{tabular}{|c|c|c|c|}
\hline
1 & 2 & 3 & 4   \\
\hline
\underline{a} & \underline{d} & a & d \\
d & b & c & b \\
b & a & \underline{b} & \underline{c} \\
c & c & d & a \\
\hline
\end{tabular}
\quad
\begin{tabular}{|c|c|c|c|}
\hline
1 & 2 & 3 & 4   \\
\hline
a & \underline{d} & \underline{a} & d \\
d & b & c & \underline{b} \\
b & a & b & c \\
\underline{c} & c & d & a \\
\hline
\end{tabular}
\quad
\begin{tabular}{|c|c|c|c|}
\hline
1 & 2 & 3 & 4   \\
\hline
a & d & \underline{a} & \underline{d} \\
d & b & c & b \\
\underline{b} & a & b & c \\
c & \underline{c} & d & a \\
\hline
\end{tabular}
\quad
\begin{tabular}{|c|c|c|c|}
\hline
1 & 2 & 3 & 4   \\
\hline
a & d & \underline{a} & d \\
\underline{d} & \underline{b} & c & b \\
b & a & b & \underline{c} \\
c & c & d & a \\
\hline
\end{tabular}
\end{center}
}

\medskip

We begin with matching $\mu_1=(1a,2d,3b,4c)$. If agents 1,3,4 meet and decide to exchange their houses by majority we can get matching $\mu_2=(1c,2d,3a,4b)$. Next, if agents 1,2,4 meet and do the same, we can get matching $\mu_3=(1b,2c,3a,4d)$, which can be again changed by majority to matching $\mu_4=(1d,2b,3a,4c)$.\footnote{Agents 1,2,4 have the same preferences over houses $b,c,d$ that they own and we get a Condorcet cycle, which can also infinitely repeat itself.} Finally, if agents 1,2,3 meet, we again can get matching $\mu_1$.

Note that in this example there is a (local) popular exchange connecting $\mu_1$ and popular matching $\mu=(1a,2b,3c,4d)$ from Table~2. In fact, as Theorem~3 demonstrates such a path exists in any instance.

\section{Most Popular and Minimal Envy Matchings}

In this section we generalize the concept of popularity to the cases where there is no popular matching. We then show that each minimal envy matching is also most popular and provide an efficient algorithm to find such matchings.

First we define popularity among a subset of agents. For a problem $(I,H,\succ)$ and a subset of agents $J\subset I$, we say that a matching $\mu$ is \textbf{popular among $\mathbf{J}$} if there does not exist another matching $\mu'$ such that $\mu'$ is preferred over $\mu$ by majority within this subset of agents, $PC_J(\mu',\mu)>PC_J(\mu,\mu')$. In other words, (reduced) matching $\mu$ is popular in the reduced problem $(J,H,\succ_{J,H})$.

We say that matching $\mu$ is \textbf{most popular} if it is popular among the largest subset $J\subset I$. Formally, we require that for each other matching $\Tilde{\mu}$ that is popular among some subset of agents $\Tilde{J}\subset I$ it has to be that set $J$ is at least as large as $\Tilde{J}$: $|J|\geq |\Tilde{J}|$. 

For an illustration consider Example~3 below.

\medskip

\textbf{Example 3.} Let there be four agents $I=\{1,2,3,4\}$ that have the following preferences (Table~6) over four houses $H=\{a,b,c,d\}$:

{
\singlespacing
\begin{center}
Table 6. Preference profile \\* \vskip 2mm
\begin{tabular}{|c|c|c|c|}
\hline
1 & 2 & 3 & 4   \\
\hline
a & b & b & b \\
b & a & c & c \\
c & c & a & a \\
d & d & d & d \\
\hline
\end{tabular}
\end{center}
}
\medskip

In Example 3, each minimal envy matching gives house $a$ to agent 1 as he is the only first agent for house $a$. The other first house~$b$, the second house~$c$ and a bad house $d$ are distributed arbitrarily among agents 2,3,4, and the owner of $d$ has envy towards the owner of~$c$. (House $d$ does not need to be matched, although this outcome would be Pareto inefficient.)

The next Observation 2 is a simple characterization of minimal envy matchings. The proof is obvious, because conditions (1) and (2) in the definition of minimal envy matching are equivalent to the corresponding conditions in Observation 2.

\medskip

\noindent{\bf Observation~2.} For a problem $(I,H,\succ)$, a matching $\mu$ has minimal envy if and only if the following two conditions hold:
\begin{enumerate}
    \item Each first house $f\in FH$ is matched to one of its first agents, $\mu(f)\in FA(f)$, and among such matchings also
    \item $\mu$ maximizes the number of agents who get their first house or their second house. 
\end{enumerate}

\medskip


Observation~2 directly implies Observation 1.

The other way to view minimal envy matchings is that these matchings assign the smallest number of bad houses -- as it maximizes the number of agents who get their first or second houses. The same can be said about most popular matchings -- the agents that are removed from consideration are precisely the agents who receive bad houses. 

However, minimal envy matchings and most popular matchings are not equivalent. Example 3 demonstrates that some most popular matchings do not satisfy minimal envy. A matching $(1d,2a,3b,4c)$ is popular among agents $\{2,3,4\}$, which makes it (one of) the most popular matchings. However, this matching does not minimize inevitable envy as agent 1 does not get house~$a$.

We formally state the relation between most popular and minimal envy matchings. The proof is in the appendix.

\medskip

\noindent \textbf{Theorem 4}. Each minimal envy matching is most popular.

\medskip

Notice however, that our definition of minimal envy matchings and most popular matchings is silent about the agents that receive bad houses. The algorithm that we provide next makes sure that the remaining houses are iteratively matched among the remaining agents in the minimal envy way. 

Our algorithm is a generalization of the original algorithm for finding popular matchings in \citep{Abraham07}. As in the original algorithm we start with a bipartite graph where agents and houses are the vertices, and each agent-vertex is connected with his first house and his second house. As in \citep{Abraham07}, first we iteratively remove leaves in this graph: we match \textit{all} single pairs of a first house and its first agent (Step 1, when there is no competition for a particular first house), then we match \textit{some} single pair of a second house and its second agent (Step 2, when there is no competition for a particular second house). After this we might get a new single pair of a first house and its first agent, and so we need to iterate from the beginning (Step 1 again).

Eventually we arrive to a bipartite graph where each house has at least two edges, i.e. no more leaves. A popular matching exists if and only if each house has two edges. Otherwise, the original procedure by \cite{Abraham07} stops here and reports that the problem does not admit a popular matching. Instead, we propose a simple extension to the algorithm. 

We select an arbitrary agent who points at some house that has three or more edges (Step~3). Then we remove this agent from the graph and iterate from the beginning: return to Step 1 to check if we obtained any new leaves in the graph, eliminate these leaves and so on until each remaining house has exactly two edges. Then each remaining agent is matched to his first house or his second house (Step~4), and for the unmatched agents we repeat the same procedure for the reduced problem with the unmatched houses (Step~5).




\medskip

{
\singlespacing
\noindent {\bf Minimal Envy Matchings (MEM)} \\
\textit{Input}: $(I,H,\succ)$; If $I=\emptyset$, then Stop;

\hspace{1em} Set matched and unmatched agents and houses: $MI=\emptyset, UI=I, MH=\emptyset, UH=H$;\\

\hspace{1em} 1: Matching of all single pairs of a first house and its first agent. 

\hspace{2em} For each $i\in UI$ such that $|FA(FH(i))\cap UI|=1$, 

\hspace{3em} Match $i$ and $FH(i)$; Remove $i$ from $UI$ to $MI$; Remove $FH(i)$ from $UH$ to $MH$;\\

\hspace{1em} 2: Matching of a single pair of a second house and its second agent. 

\hspace{2em} If some $i\in UI$ have $|SA(SH(i))\cap UI|=1$,  

\hspace{3em} Match $i$ and $SH(i)$; Remove $i$ from $UI$ to $MI$; Remove $SH(i)$ from $UH$ to $MH$; 

\hspace{3em} Go to Step 1;\\
  
\hspace{1em} 3: Excluding of some agent.
    
\hspace{2em}    If some $i\in UI$ have $|FA(FH(i))\cap UI|> 2$ or $|SA(SH(i))\cap UI|> 2$, 
    
\hspace{3em}    Remove $i$ from $UI$; Go to Step 1;\\
 
\hspace{1em} 4: Matching cycles.
   
\hspace{2em} While $UI\neq \emptyset$,

\hspace{3em} Choose some $i\in UI$;

\hspace{3em} While $i\in UI$,

\hspace{4em} Match $i$ and $FH(i)$; Remove $i$ from $UI$ to $MI$; 

\hspace{4em} Remove $FH(i)$ from $UH$ to $MH$;

\hspace{4em} Match $j\in FA(FH(i))\cap UI$ and $SH(j)$; Remove $j$ from $UI$ to $MI$; 

\hspace{4em} Remove $SH(j)$ from $UH$ to $MH$; 

\hspace{4em} If $SA(SH(j))\cap UI\neq \emptyset$, choose $i\in SA(SH(j))\cap UI$;

\hspace{1em} 5: Run the algorithm on Input $(I\setminus MI,H \setminus MH,\succ_{I\setminus MI,H \setminus MH})$.
}

\medskip

Now we can present the main result of this section: the MEM algorithm produces a minimal envy matching that is also Pareto efficient. The proof is in the appendix.

\medskip

\noindent{\bf Theorem 5.} The induced output of the MEM algorithm
\begin{itemize}
    \item[i)] is a minimal envy matching,
    \item[ii)] is Pareto efficient,
    \item[iii)] is obtained in cubic time in the number of agents $|I|$.
\end{itemize}

\section{Conclusions}

In the current paper we study fair matchings defined as popular matchings or, alternatively, minimal envy matchings. We propose a characterization of ``global'' popularity via ``local'' popularity, and also show that
local majority-based exchanges lead to a globally popular matching. 

We also constructively extend the definition of popularity to the cases where a popular matching does not need to exist. In general, a matching is most popular if it is popular among the largest number of agents. We show that each minimal envy matching is a most popular matching, and provide an efficient algorithm to find these matchings.

There might be other reasonable alternative extensions of popular matchings adopted from voting theory, for instance Schwartz set. The other way, using the language of minimal envy, one can find matchings that eliminate envy among the group of disadvantaged agents. For example, imagine that in some most popular matching the agents that received bad houses do not envy each other (as each of them received a ``third`` house -- the best house after all first houses and second houses are removed). This matching intuitively seems to be more fair than some other most popular matching where this is not the case. However, we conjecture that finding such matching is NP-hard.

One open question is about the convergence speed of decentralized
popular markets. To answer this question one may need to design a
more efficient algorithm: our greedy algorithm does many unnecessary
steps, for instance when it repeatedly runs the same chains. We cannot simply avoid these steps as then we cannot build a triple that blocks the current matching. However, it might be possible using alternative algorithms.

Another open question is about popular markets in instances when popular matchings do not exist. Perhaps, these markets converge to some stationary probabilistic distribution over the set of matchings, and it is reasonable to deem the more probable matchings
as more popular. Both questions are interesting but hard.

Our results also raise practically relevant questions for market designers: how can minimal envy and most popular matchings be found in relevant matching markets. For instance, in the centralized school choice market where schools have coarse priorities, one might want to find the minimal envy matching among all stable matchings. This way the students will have less envy towards the students of the same priority category than otherwise. In the same time the final matching (and the underlying tie-breaking that supports this matching) is fair from the social perspective since it is the most popular among all stable matchings. Such refinements of tie-breaking will also have positive efficiency implications akin the results by \cite{erdil2008s}.

A fundamental question is to develop alternative principles of fairness applicable for the indivisible object allocation setting considered here. Despite the fact that the setting precludes many standard principles of fairness like symmetry and fair-share-guaranteed, there might be other reasonable approaches to fairness that do not rely on envy or popularity.

\bigskip


{\bf\large APPENDIX}

\medskip

\textsl{Proof of Theorem 2}.
The ''only if'' part is straightforward: each popular matching $\mu$ is locally popular. For a contradiction, assume that there is a triple of agents $i,j,k\in I$ and another matching $\mu'$ same as $\mu$ for all other agents and such that it is preferred over $\mu$: $PC_{\{i,j,k\}}(\mu',\mu)>PC_{\{i,j,k\}}(\mu,\mu')$. Then $\mu$ cannot be popular among all agents since all other agents are indifferent and thus:
\begin{equation*}
    PC_I(\mu',\mu)-PC_I(\mu,\mu')=PC_{\{i,j,k\}}(\mu',\mu)-PC_{\{i,j,k\}}(\mu,\mu')>0.
\end{equation*}

The ''if'' part we also prove by contradiction. For a contradiction, assume that there is a matching $\mu$ that is locally popular, but it loses in pairwise comparison to some other matching $\mu'$: $PC_I(\mu',\mu)>PC_I(\mu,\mu')$. Consider all agents that have different houses in these two matchings, denote the set of these agents as $I_1 = \{ i\in I : \mu(i) \neq \mu'(i) \}$. (In what follows we will change the notation of these agents for convenience).

We partition all agents into those who participate in a trading cycle, i.e. exchange their matched houses among themselves, and a trading chain, i.e. those that are matched in $\mu'$ to a previously empty house or whose house in $\mu$ becomes empty in~$\mu'$.

We first deal with chains. Consider an arbitrary agent $j_1\in I_1$ that received a previously empty house $\mu'(j_1) \in H\setminus \mu(I)$, $\mu(\mu'(j_1))=\emptyset$. If $j_1$'s house is empty in $\mu'$, $\mu'(\mu(j_1))=\emptyset$, then we get a chain of size 1. Otherwise there is some agent $j_2$ such that $\mu'(j_2)=\mu(j_1)$. If $j_2$'s house is empty in $\mu'$, $\mu'(\mu(j_2))=\emptyset$, then we get a chain of size 2. Otherwise, we continue in the same way until we find the last agent in the chain. Similarly, determine chains for each agent that receives a previously empty house. Denote the set of agents participating in a chain as $J_1$.

We then deal with cycles. Consider an arbitrary agent not from any chain $i_1\in I_1\setminus J_1$, $\mu(i_1) \neq \mu'(i_1)$. Consider agent $i_2$ that owns house $\mu'(i_1)$, $i_2 = \mu(\mu'(i_1))$.
Agent $i_2$ also does not belong to any chain, $i_2\in I_1\setminus J_1$ and as $\mu(i_2) = \mu'(i_1)$, then $i_2\neq i_1$. If the two agents just exchanged their houses, $\mu'(i_2) = \mu(i_1)$, then we get a trading cycle $(\mu(i_1), i_1, \mu'(i_1), i_2)$ of length 2. Otherwise, if $\mu'(i_2) \neq \mu(i_1)$, then consider agent $i_3 = \mu(\mu'(i_2))$. Since $\mu(i_3) = \mu'(i_2) \neq \mu(i_1)$, then $i_2\neq i_3$, $i_1\neq i_3$ and $i_3\in I_1\setminus J_1$.

And so forth until we get a cycle of length at least 2 and at most $|I_1\setminus J_1|$. In the same way we find all trading cycles among all other agents.

Thus, the set $I_1$ and the set of corresponding houses $\mu(I_1)\cup \mu'(I_1)$ is partitioned into trading chains of size at least 1 and cycles of size at least 2.

By assumption $PC_I(\mu',\mu)>PC_I(\mu,\mu')$, there is at least one trading chain or one trading cycle such that more than half of its agents prefer $\mu'$ over $\mu$. Formally, if $I_{TC}$ denotes the set of agents in this chain or cycle, $PC_{I_{TC}}(\mu',\mu)>PC_{I_{TC}}(\mu,\mu')$.

If $I_{TC}$ form a cycle, then we can find two neighbouring agents $i,j\in I_{TC}$, $j=\mu(\mu'(i))$, that both prefer $\mu'$ over $\mu$. If this trading cycle is of length 2, then consider a new matching $\mu''$ that is identical to $\mu$ for each agent except $\{i,j\}$ and same as $\mu'$ for these pair, $\mu''(i)=\mu'(i)$, $\mu''(j)=\mu'(j)$. Then by adding one other arbitrary agent we get a triple of agents that prefer $\mu''$ over $\mu$ by majority -- contrary to our premise. If this trading cycle is of length more than 2, then consider the next neighbouring agent $k=\mu(\mu'(j))$. Consider now a new matching $\mu''$ that is identical to $\mu$ for each agent except $\{i,j,k\}$ and $\mu''(i)=\mu'(i)$, $\mu''(j)=\mu'(j)$, and $\mu''(k)=\mu(i)$. The triple of agents $i,j,k$ prefers $\mu''$ over $\mu$ by majority: $PC_{\{i,j,k\}}(\mu'',\mu)>PC_{\{i,j,k\}}(\mu,\mu'')$, contrary to our premise.

If $I_{TC}$ forms a chain of length 1, $I_{TC}=\{i_1\}$, then consider a new matching $\mu''$ constructed as before: $\mu''$ is identical to $\mu$ for each agent except for $i_1$, $\mu''(i_1)=\mu'(i_1)$. A triple of agents $i_1$ and two arbitrary agents $i_2$, $i_3$ prefers $\mu''$ over the original matching $\mu$: $PC_{\{i_1,i_2,i_3\}}(\mu'',\mu)>PC_{\{i_1,i_2,i_3\}}(\mu,\mu'')$, contrary to our premise.

If $I_{TC}$ forms a chain of length 2, then both agents in $I_{TC}$ are better off in $\mu'$ compared to $\mu$. By adding one other arbitrary agent we get a triple of agents that prefers a similarly constructed $\mu''$ over $\mu$ by majority, contrary to our premise.

If the length of the chain is above 2, then either (1) we can find two neighbouring agents $i,j\in I_{TC}$, $j=\mu(\mu'(i))$, that both prefer $\mu'$ over $\mu$, or (2) the chain begins and ends with agents that are better off in $\mu'$ compared to $\mu$ (and agents in between interchange). In case (1) we take the triple of these agents $i,j$ and the previous owner of $j$'s house $k=\mu(\mu'(j))$ (if $j$'s house was empty, then take an arbitrary $k$). This triple $i,j,k$ prefers a similarly constructed $\mu''$ over $\mu$ by majority, contrary to our premise.

In case (2) we take the triple of agents as the first agent in the chain $j_1$, $\mu(\mu'(j_1))=\emptyset$, the last agent $j_k$, $\mu'(\mu(j_k))=\emptyset$, and the one before the last $j_{k-1}$. The triple $j_1, j_{k-1},j_k$ prefers a similarly constructed $\mu''$ over $\mu$ by majority, contrary to our premise. \hfill
$\blacksquare$

\medskip

\textsl{Proof of Theorem 3}.

The first part of the algorithm.

Let $\mu=\mu_0$ be the arbitrary initial matching where each agent is endowed with some house $h$ or $\emptyset$. Let us fix some ordering of agents $I=\{i_1,\ldots,i_n\}$.
Denote the current matchings by $\mu_j, j=0,\ldots,l$. Each next matching $\mu_{j+1}$ is the same as previous matching $\mu_j$ except some local popular exchange. Denote the number of agents who get a bad house by $\beta(\mu_j)$. Note that $n-\beta(\mu_j)$ agents get either a first house or a second house.

For steps $k=1,\ldots, n$ we make the following local popular exchanges.

If in step $k$ house $\mu_j(i_k)$ is the best house for agent $i_k$, then proceed to step $k+1$ without changing the current matching $\mu_j$. Otherwise, consider house $h\neq \mu_j(i_k)$ that is the best house of agent $i_k$. If this house $h$ is empty, $\mu_j(h)=\emptyset$, then we give it to agent $i_k$ in the new matching, $\mu_{j+1}(i_k)=h, \mu_{j+1}(h)=i_k$. Otherwise, consider the owner of $h$, $\mu_j(h)$.

If $h$ is the best for its owner $\mu_j(h)$, then proceed to the next step $k+1$ without changing the current matching $\mu_j$. Otherwise, consider the best house for agent $\mu_j(h)$: $h'\neq h$. If $h'=\mu_j(i_k)$ or $\mu_j(h')=\emptyset$ then make the mutually beneficial two-way exchange: $\mu_{j+1}(i_k)=h, \mu_{j+1}(h)=i_k, \mu_{j+1}(\mu_j(h))=h', \mu_{j+1}(h')=\mu_j(h)$. Otherwise, if $\mu_j(h')\notin\{i_k, \mu_j(h), \emptyset\}$ we make the three-way exchange:  $\mu_{j+1}(i_k)=h, \mu_{j+1}(h)=i_k, \mu_{j+1}(\mu_j(h))=h', \mu_{j+1}(h')=\mu_j(h), \mu_{j+1}(\mu_j(h'))=\mu_j(i_k), \mu_{j+1}(\mu_j(i_k))=\mu_j(h')$. This exchange is beneficial for at least two of the three agents.

After each of the above exchanges the number of agents that own their best houses goes up, and each agent gets his best house unless it is taken by some other agent. Thus after $x\leq n$ local popular exchanges we get a new matching $\mu_x$ where each agent gets either his first house, his second house, or a bad house. At least $\max\{x,1\}$ agents get their first house, therefore $\beta(\mu_x)\leq \min\{n-x, n-1\}$. 

\medskip

The second part of the algorithm.

We will make exchanges that \emph{weakly} decrease the number of agents with a bad house, $\beta(\mu_j)\geq \beta(\mu_{j+1})$.

Consider some agent $\mu_j(t)$ that gets a bad house $t$. If his second house $s$ is free, $\mu_j(s)=\emptyset$, we give him $s$: $\mu_{j+1}(\mu_j(t))=s, \mu_{j+1}(s)=\mu_j(t)$ and decrease $\beta(\mu_j)$ by one, $\beta(\mu_{j+1})= \beta(\mu_j)-1$. Otherwise there is some agent $\mu_j(s)$ that owns $s$, and $s$ might be his bad house or his second house (but not his first house from the definition of second house). We now study these two cases.

1. Let $s$ be a bad house for $\mu_j(s)$. Denote the second house of $\mu_j(s)$ as $h$. If $h=t$ or $h$ is empty, $\mu_j(h)=\emptyset$, then make the two-way exchange decreasing $\beta(\mu_j)$ by two, $\beta(\mu_{j+1})= \beta(\mu_j)-2$: $\mu_{j+1}(\mu_j(t))=s, \mu_{j+1}(s)=\mu_j(t),\mu_{j+1}(\mu_j(s))=h, \mu_{j+1}(h)=\mu_j(s)$. Otherwise, make the three-way exchange $\mu_{j+1}(\mu_j(t))=s, \mu_{j+1}(s)=\mu_j(t),\mu_{j+1}(\mu_j(s))=h, \mu_{j+1}(h)=\mu_j(s), \mu_{j+1}(\mu_j(h))=t, \mu_{j+1}(t)=\mu_j(h)$, decreasing $\beta(\mu_j)$ by 1, 2 or 3 depending on how agent $\mu_j(h)$ ranks house~$t$.

2. Let $s$ be the second house for $\mu_j(s)$. Let $f$ be the first house for agent $\mu_j(s)$. From the first part of the algorithm we know that $f$ is also the first house of his owner $\mu_j(f)$. Make the following three-way exchange: $\mu_{j+1}(\mu_j(t))=s, \mu_{j+1}(s)=\mu_j(t),\mu_{j+1}(\mu_j(s))=f, \mu_{j+1}(f)=\mu_j(s), \mu_{j+1}(\mu_j(f))=t, \mu_{j+1}(t)=\mu_j(f)$. If $t$ is the second house for agent $\mu_j(f)$, then $\beta(\mu_j)$ decreases by one, $\beta(\mu_{j+1})= \beta(\mu_j)-1$.

Thus $\beta(\mu_j)$ is only constant, $\beta(\mu_{j+1})= \beta(\mu_j)$, if house $s$ is the second house for both $\mu_j(t)$ and $\mu_j(s)$, house $f$ is the first house for both $\mu_j(s)$ and $\mu_j(f)$, and house $t$ is a bad house for both agents $\mu_j(t)$ and $\mu_j(f)$. Denote such exchange as \emph{bad} (see Table~7). We show now that a sequence of these bad exchanges in which $\beta(\mu_j)$ remains constant is finite.

\medskip

{
\singlespacing
\begin{center}
Table 7. Current matchings $\mu_j, \mu_{j+1}$ before and after a bad three-way exchange which keeps $\beta(\mu_j)=\beta(\mu_{j+1})$ \\* \vskip 2mm
\begin{tabular}{|c|c|c|c|}
\hline
house & $\mu_j(t)$ & $\mu_j(s)$ & $\mu_j(f)$   \\
\hline
first &  & f & \underline{f}   \\
second & s & \underline{s} &     \\
bad & \underline{t} &   & t    \\
\hline
\end{tabular}
\quad
\begin{tabular}{|c|c|c|}
\hline
$\mu_j(t)$ & $\mu_j(s)$ & $\mu_j(f)$   \\
\hline
  & \underline{f} & f   \\
\underline{s} & s &     \\
t &   & \underline{t}   \\
\hline
\end{tabular}
\end{center}
}

\medskip

2.1 Let $f$ be the first house also for agent $\mu_j(t)$. For convenience denote $f=f_1, s=s_1, \mu_j(t)=1, \mu_j(s)=2, \mu_j(f)=3$. By Hall's theorem the second house for agent 3 cannot be the same as $s_1$, $s_3 \neq s_1$ (otherwise three agents have the same first house and the same second house, and thus a popular matching does not exist). After the bad exchange among agents 1,2,3 the bad house $t$ is matched to agent 3. Consider another chain of three agents that starts with the bad house $t$. Denote $\mu_j(s_3)=4$. Note that $f_4 \neq f_1$ (otherwise four agents have the same first house, two of them have the same second house, and the other two of them also have the same second house, and thus a popular matching does not exist). Denote $\mu_j(f_4)=5$. By Hall's theorem $s_5\notin \{ s_1, s_3 \}$ (otherwise, similar to the previous arguments the popular matching does not exist). After the bad exchange between agents 3,4,5 the bad house is matched with agent 5 (see Table~8), and so forth.

\medskip

{
\singlespacing
\begin{center}
Table 8. Current matchings $\mu_j, \mu_{j+2}$ before and after two bad three-way exchanges which keep $\beta(\mu_j)=\beta(\mu_{j+2})$ \\* \vskip 2mm
\begin{tabular}{|c|c|c|c|c|c|}
\hline
house & 1 & 2 & 3 & 4 & 5   \\
\hline
first & $f_1$ & $f_1$ & $\underline{f_1}$ & $f_4$ & $\underline{f_4}$ \\
second & $s_1$ & $\underline{s_1}$ & $s_3$ & $\underline{s_3}$ & $s_5$ \\
bad & \underline{t}     &       & t     &       & t    \\
\hline
\end{tabular}
\quad
\begin{tabular}{|c|c|c|c|c|}
\hline
1 & 2 & 3 & 4 & 5   \\
\hline
$f_1$ & $\underline{f_1}$ & $f_1$ & $\underline{f_4}$ & $f_4$ \\
$\underline{s_1}$ & $s_1$ & $\underline{s_3}$ & $s_3$ & $s_5$ \\
t     &       & t     &       & \underline{t}    \\
\hline
\end{tabular}
\end{center}
}

\medskip

Note that in this case $\beta(\mu_j)\leq n-2$. In each such bad exchange two new agents enter the chain, these agents own their first and second houses. Then, we need not more than $(n-\beta(\mu_j))/2$ bad exchanges and one additional exchange to reduce~$\beta(\mu_j)$. Hence, the total number of local popular exchanges reducing $\beta(\mu_j)$ is not more than
\begin{equation*}
\frac{n-\beta(\mu_j)}{2}+1 \leq n-\beta(\mu_j).
\end{equation*}

2.2 Let the first house $f_1$ for agent $\mu_j(t)$ be different from house $f$. Denote $f=f_2, s=s_1, \mu_j(t)=1, \mu_j(s)=2, \mu_j(f)=3$. After one bad exchange agent 3 would be matched to house~$t$.

Assume that the second house for agent 3 $s_3\neq s_1$ -- we did not meet $s_3$ earlier in the chain. Consider agent 4 that owns his second house $s_3$. Assume that agent 4's first house $f_4$ was note previously in the chain: $f_4\neq f_1,f_2$.

Consider the next agent 5 and so on: we get a chain of agents such that each two neighbours have either the same first house or the same second house. Eventually we arrive to some agent $k$ that has the same first or second house as earlier in the chain.

Let agent $k$ be the first agent in the chain such that his first house has already appeared in the chain. In this case $k$ is even. Then after $(k-2)/2$ bad exchanges in one direction agent $k-1$ gets the bad house $t$. Then agent $k-1$ reverses the direction of bad exchanges such that agent~$k+1$ gets the bad house $t$, which happens after not more than than $k/2$ bad exchanges. And we see that not less than $k$ agents get their first or second houses, $k\leq n-\beta(\mu_j)$.

Now agent $k+1$ continues the bad exchanges. By Hall's theorem, in each such bad exchange two new agents enter the chain, these agents own their first and second houses. Then, we need not more than $(n-\beta(\mu_j) -k)/2$ bad exchanges and one additional exchange to reduce~$\beta(\mu_j)$. Hence, for even $k$ the total number of exchanges reducing $\beta(\mu_j)$ is not more than
\begin{equation*}
\frac{k-2}{2}+\frac{k}{2}+\frac{n-\beta(\mu_j) -k}{2}+1 =
\frac{n-\beta(\mu_j)}{2}+\frac{k}{2} \leq n-\beta(\mu_j).
\end{equation*}

Similarly, let agent $k$ be the first agent in the chain such that his second house has already appeared in the chain. In this case $k$ is odd. Then after $(k-1)/2$ bad exchanges in one direction agent $k$ gets the bad house $t$. Then agent $k$ reverses the direction of bad exchanges such that agent 1 gets house $f_1$ and agent~$k+1$ gets the bad house $t$, which happens after not more than than $(k-1)/2$ bad exchanges. And we see that not less than $k$ agents get their first or second houses, $k\leq n-\beta(\mu_j)$.

Now agent $k+1$ continues the bad exchanges. By Hall's theorem, in each such bad exchange two new agents enter the chain, these agents own their first and second houses. Then, we need not more than $(n-\beta(\mu_j) -k)/2$ bad exchanges and one additional exchange to reduce~$\beta(\mu_j)$. Hence, for odd $k$ the total number of local popular exchanges reducing $\beta(\mu_j)$ is not more than
\begin{equation*}
\frac{k-1}{2}+\frac{k-1}{2}+\frac{n-\beta(\mu_j) -k}{2}+1 =\frac{n-\beta(\mu_j)}{2}+\frac{k}{2} \leq n-\beta(\mu_j).
\end{equation*}

Eventually, in the second part of the algorithm after at most $n-\beta(\mu_j)$ local popular exchanges we decrease $\beta(\mu_j)$. In the worst case $n-1$ agents have a bad house after the first part of the algorithm, therefore, including the first part of the algorithm the upper bound is $1+1+2+\ldots+(n-1)=(n^2-n+2)/2$. \hfill $\blacksquare$





\medskip

\textsl{Proof of Theorem 4}.
For a contradiction, consider a minimal envy matching $\mu$ that is not most popular: there exists another matching $\mu'$ popular among a larger set $J'$, i.e. for each set $J$ such that $\mu$ is popular among $J$, $|J'|>|J|$. (In particular, the latter is true for $J$ consisting of agents that get their first or second houses in $\mu$.)

We now modify $\mu'$ such that a resulting matching has less envy than $\mu$. Consider an arbitrary house $h$ that is matched in $\mu'$ to some agent $i\in J'$. As $\mu'$ is popular among $J'$, house $h$ is either $i$'s first house in the reduced problem with set $J'$ (and thus $h$ is also $i$'s first house in the original problem with set $I$), or $h$ is $i$'s second house in the reduced problem $(J',H,\succ_{J',H})$. Consider the latter case. 

If $h$ is NOT $i$'s second house in the original problem, then there exists agent $j\in I \setminus J'$ for whom $h$ is a first house, $h=FH(j)$. Match $h$ to $j$, and give agent $i$ its most preferred unmatched house. Notice that the number of first and second houses is at least as large as in $\mu'$. Do the same for each such house $h$ (matched to an agent in $J'$ that is its second agent in the reduced problem while its first agent is in $I\setminus J'$). 

As a result we match each first house to one of its first agents (either in $J'$ as already done in $\mu'$ or in $I\setminus J'$ by modifying $\mu'$), and the total number of agents who get their first or second houses is at least $|J'|$ and hence larger than in $\mu$. Therefore, $\mu$ is not minimal envy matching.
\hfill $\blacksquare$

\medskip

\noindent \textit{Proof of Theorem 5}. 

Statement i) Let the MEM algorithm produce matching $\mu$. Consider some matching $\mu'$ that is a minimal envy matching. Both matchings $\mu, \mu'$ match each first house to one of its first agents. We show that $\mu$ and $\mu'$ match the same number of agents to their first or second houses and therefore $\mu$ is a minimal envy matching.

Denote $I_1$ the set of agents that are matched until we reach Step 3, i.e., while we only match leaves and no agent is excluded. Each agent in $I_1$ gets his first or second house in $\mu$: for each $i\in I_1$ we have $\mu(i)\in FH(i)\cup SH(i)$. We first show that each agent in $I_1$ also gets his first or second house in $\mu'$. 

For a contradiction, let there be agent $i_1\in I_1: h_0 = \mu'(i_1)\notin FH(i_1)\cup SH(i_1)$. 

Since $i_1\in I_1$, in $\mu$ he received either his first or second house $h_1=\mu(i_1)\in FH(i_1)\cup SH(i_1)$. House $h_1\neq h_0$ and in $\mu'$ house $h_1$ is matched to some agent $i_2=\mu'(h_1)\notin \{\emptyset,i_1\}$ for whom $h_1$ is his first or second house, $h_1\in FH(i_2)\cup SH(i_2)$, otherwise a minimal envy matching $\mu'$ should match $h_1$ to $i_1$. Therefore, $i_2\in I_1$ since $h_1$ was matched as a leaf and thus $i_2$ was matched earlier than $i_1$ in the MEM algorithm.

Since $i_2\in I_1$, in $\mu$ he received either his first or second house $h_2=\mu(i_2)\in FH(i_2)\cup SH(i_2)$. House $h_2\notin \{h_0,h_1\}$ and in $\mu'$ house $h_2$ is matched to some agent $i_3=\mu'(h_2)\notin \{\emptyset,i_1,i_2\}$ for whom $h_2$ is his first or second house, $h_2\in FH(i_3)\cup SH(i_3)$, otherwise a minimal envy matching $\mu'$ should match $h_2$ to $i_2$ and $h_1$ to $i_1$. Therefore, $i_3\in I_1$ since $h_2$ was matched as a leaf and thus $i_3$ was matched earlier than $i_2$ in the MEM algorithm.

We get a contradiction with finiteness of $I_1$: an infinite chain of different agents $(i_1,i_2,i_3,\ldots)$ such that for each two neighbors $i_{k-1},i_k$ we have $\mu(i_{k-1})=\mu'(i_k)$ and each agent $i_k\in I_1$. 



It remains to show that $\mu'$ could not match agents in $I\setminus I_1$ `better' than $\mu$ does.

Since each agent $i\in I_1$ received a leaf in the MEM algorithm, we have that $\mu(i)$ is not a first house or second house of any agent $j\in I\setminus I_1$: $\mu(i)\notin FH(j)\cup SH(j)$. Therefore, the MEM algorithm distributes all houses valuable for \textit{each} such agent $j\in I\setminus I_1$: $\mu$ matches each first house $FH(j)$ to an agent in $I\setminus I_1$ that deems it as his first house, and each second house $SH(j)$ to an agent in $I\setminus I_1$ that deems it as his second house.


Statement ii) For a contradiction let there be a Pareto improvement over matching $\mu$ induced by the MEM algorithm, and let some set of agents $J$ strictly benefit. For all possible improvements over $\mu$ consider agent $i\in J$ who is matched by the MEM algorithm earlier than any other agent in such sets $J$. Let $i$ be matched in run $k$ of the MEM algorithm.

Which house does agent $i$ get? In the corresponding run $k$ of the MEM algorithm agent $i$ received either his current first house (i.e., his most preferred house among all houses that remain unmatched after earlier runs) or his current second house. 

If $i$ received his current first house, then he can only benefit by getting some house $h$ matched in the earlier run of the MEM algorithm, but since $i$ is the earliest matched agent who can Pareto improve, agent $\mu(h)$ is worse off. Alternatively, if $i$ received his current second house, he can additionally benefit only by getting some other agent $j$'s current first house (matched in run $k$ of the MEM algorithm). But agent $j$ can only benefit by getting some house $g$ matched in the earlier run of the MEM algorithm. Agent $\mu(g)$ is matched earlier than $i$ and thus is worse off since $i$ is the earliest matched agents who can Pareto improve.

Statement iii) Each run of the MEM algorithm has the same complexity $O(|I|^2)$ as in the original algorithm by \cite{Abraham07} and reduces the size of the problem at least by two agents and two houses.

\hfill
$\blacksquare$

\bibliographystyle{apalike}
\bibliography{popularity}

\end{document}